\documentclass[review]{elsarticle}

\usepackage{lineno,hyperref}
\usepackage{subcaption}
\usepackage{graphicx}
\usepackage{amsmath}
\usepackage{xcolor}
\usepackage{amssymb}
\usepackage{longtable}

\journal{International Journal of Heat and Mass Transfer}

\definecolor{negContour}{HTML}{430053}
\definecolor{posContour}{HTML}{fee725}

\author[stfc]{Alex Skillen\corref{ca}}
\ead{alex.skillen@stfc.ac.uk}

\author[ibm]{Ma{\l}gorzata J. Zimo{\'n}}
\author[ibm]{Robert Sawko}
\author[rr]{Ryan Tunstall}
\author[stfc]{Charles Moulinec}
\author[stfc]{David R Emerson}

\cortext[ca]{Corresponding author}

\address[stfc]{STFC Daresbury Laboratory, Warrington WA4 4AD, UK}
\address[ibm]{IBM Research UK, Daresbury Laboratory, Warrington WA4 4AD, UK}
\address[rr]{Rolls-Royce, Raynesway, Derby, DE21 7BE, UK}

\begin{document}

\begin{frontmatter}

\title{Thermal transients in a U-bend}

\begin{abstract}
    We study numerically the propagation of a hot thermal transient through a
    U-bend via an ensemble of wall-resolved large eddy simulations. Conjugate heat transfer
    between fluid and solid domains is accounted for. The flow is in a fully
    turbulent mixed convection regime, with a bulk Reynolds number of
    $10,000$, a Richardson number of $2.23$, and water as the working fluid (Prandtl number = $6$). These conditions lead to strong
    thermal stratification, with buoyancy-induced secondary flows, and the
    generation of a large and persistent recirculation region.

    The evolution of Dean vortices as the thermal transient passes is studied.
    It is found that baroclinic vorticity generation dominates over a large
    period of the transient, due to the thermal inertia of the wall.
    Gravitational buoyancy leads to a reversal of the counter-rotating vortex
    pair. The impact of this reversal on the swirl-switching and
    secondary-current losses is assessed. It is found that low frequency modes
    are suppressed in the reversed-vortex state. 
\end{abstract}


\end{frontmatter}


\section{Introduction}
As a fluid flows around a bend, a local imbalance between the centripetal force and the opposing pressure gradient acts to transport low-inertia near-wall fluid towards the
centre of curvature. This induces a counter-rotating vortex pair, known as Dean
vortices, as fluid is subsequently transported  back along the symmetry plane.
The impact these Dean vortices have upon heat and mass transfer within the pipe
is generally significant, and can be characterised by the dimensionless Dean
number: $Dn \equiv Re\sqrt{D (2R_c)^{-1}}$ (where $Re$, $D$ and $R_c$ are the bulk
Reynolds number, the pipe inner-diameter and radius of curvature,
respectively).

A number of studies have systematically quantified the behaviour of flow around bends with circular cross section. 
Improvements in numerical modelling of developing laminar flows in U-shaped pipe were suggested by Humphrey at el. \cite{humphrey1985some}, 
enabling better understanding of the secondary fluid motions. In a turbulent setting, secondary flow development
was investigated by Azzozal et al. \cite{azzola1986developing} both experimentally (using a laser Doppler anemometer) and numerically. 
These studies were continued by Baughn et al. \cite{baughn1987local} focusing on local heat transfer measurements for similar turbulent configurations. 
Although the problem of flows in curved pipes has been 
extensively studied over the past several decades, there is still a relative lack in understanding in the flow physics, and particularly the instabilities involved.

At moderate to high Dean numbers, instabilities in the flow manifest as a
cyclical change in the strength and size of the left and right Dean vortices
relative to one another \cite{carlsson2015swirl}. This phenomenon is known as
swirl-switching. A number of computational
\cite{carlsson2015swirl,rutten2005large} and experimental
\cite{sakakibara2012measurement,hellstrom2013turbulent,kalpakli2013turbulent}
studies have investigated the physics of swirl-switching, and a review has been
conducted by Vester et al. \cite{vester2015pod}. Transverse forces induced by
swirl-switching can have significant impact upon the fatigue-life of plant
components \cite{tunstall2016large}. It is generally accepted the switching
occurs as a gradual low-frequency shift between states, rather than an abrupt
change from one bi-stable state to another \cite{rohrig2015comparative}.

A range of frequencies associated with swirl-switching have been reported.
Although reported values vary (often significantly) between different studies,
it is generally accepted the swirl-switching has both low and high frequency
modes. For example, Hellstr\"om et al. \cite{hellstrom2013turbulent} report two
characteristic Strouhal numbers, $St$, of $0.16$ and $0.33$ ($St \equiv f D
U^{-1}$, where $U$ is the bulk velocity and $f$ is the frequency). Recent work by
Wang et al. \cite{wang2018direct} has highlighted that the recorded frequency
of swirl-switching is highly sensitive to local flow properties, which may
explain the discrepancies in recorded frequencies between different authors.
Furthermore, their study highlights the role of large upstream structures on
the low-frequency mode, while the higher-frequency mode originates from
structures generated within the pipe-bend.

In the present study, we are concerned with the dynamics of flow in a U-bend as
a hot thermal transient propagates throughout the domain. This case is relevant to the loop seal of a pressurised water nuclear fission reactor, amongst others. Both hot and cold
thermal transients in a U-bend have been studied experimentally by Viollet
\cite{viollet1987observation} by applying a linear ramp to the inlet
temperature over a short duration.  Viollet observed that under the conditions
of sufficiently low Reynolds number and Froude number, thermal stratification
tends to occur. Stratification can significantly alter the flow, leading to the
formation of a large buoyancy-induced recirculation region and steep temperature gradients.
Furthermore, cyclical changes in inlet temperature, alternating between hot and cold states, can lead to thermal fatigue. 

The impact of buoyancy upon Dean vortices has been investigated in a number of
studies. Lingrani et al. \cite{ligrani1996effects} studied experimentally the
effect Dean vortices have upon surface heat transfer in a curved channel. They
were primarily interested in forced convection flows, but faced challenges in
removing buoyancy effects from their experiments. Mixed convection Nusselt
numbers were therefore measured, and a correlation was given to convert mixed
convection Nusselt numbers into forced convections ones.

Ciofalo et al. \cite{ciofalo2015influence} investigated computationally the
influence of both gravitational and centrifugal buoyancy for laminar flow within a
coiled pipe. They performed a parameter study over a range of Richardson numbers
with a linear increase in the pipe-wall temperature in the axial direction. 

Kurnia et al. \cite{kurnia2016numerical} performed calculations of laminar flow
in straight and helical pipes of various cross section. They fixed the wall
temperature and considered three different temperature differentials between
wall and inlet. The dilatable working fluid (air) lead to conditions in which
both gravitational and centrifugal buoyancy effects were significant. The study
highlighted the formation of secondary vortices due to buoyancy in straight
pipes, as well as the interaction between buoyancy-driven secondary flows and
Dean vortices in the helical pipe cases.  

In the aforementioned studies, \cite{ligrani1996effects, ciofalo2015influence,
kurnia2016numerical}, the temperature differential between the near-wall fluid
and bulk-fluid was generated by heating (or cooling) the wall. To the best of
our knowledge, the impact a \emph{thermal transient} has upon Dean vortex
dynamics and swirl-switching has not been previously studied. It is anticipated
that a thermal transient, in conjunction with high wall thermal inertia, would
lead to conditions in which large radial temperature (and density) gradients are
present. In \emph{mixed convection} flows, the gravitational and centrifugal
buoyancy due to these density gradients has the potential to generate secondary
flow, similar to those observed by Kurnia et al. \cite{kurnia2016numerical}.
The aim of this work is to test this hypothesis, and assess the impact a hot
thermal transient has upon Dean vortex dynamics and swirl-switching within a
U-bend configuration under turbulent mixed convection conditions. 

This paper is structured as follows. In Section \ref{sec:methodology} we
outline our methodology. In Section \ref{sec:results}, key findings are presented. Finally,
in Section \ref{sec:conclusion} conclusions are drawn, and recommendations
for future work are made.

\section{Methodology}\label{sec:methodology}

\subsection{Governing Equations}\label{sec:gov}
We employ an operator splitting strategy to decompose the problem into fluid
and solid domains. The fluid flow is governed by the incompressible filtered
Navier-Stokes equations (i.e. Large Eddy Simulations, (LES)) with the Boussinesq
approximation to account for buoyancy (the impact of this approximation is to be discussed in Section \ref{sec:studyParam}), and a reduced filtered-energy equation
to account for heat transfer:

\begin{equation}
	\label{eq:LEScon}
	\frac{\partial \overline{u}_i}{\partial x_i}=0
\end{equation}

\begin{equation}
	\label{eq:LESns}
    \frac{\partial \overline{u}_i}{\partial t}
    +
    \frac{\partial (\overline{u}_i \overline{u}_j)}{\partial x_j}
    =
    \frac{\partial}{\partial x_j} \left[ -\frac{\overline{p}}{\rho_0}\delta_{ij}
    +
  \nu \left(
    \frac{\partial \overline{u}_i}{\partial x_j}
    +
    \frac{\partial \overline{u}_j}{\partial x_i}
  \right)
  - \tau_{ij} \right] - \rho_0 \beta \left(\overline{T} - T_0 \right) g_i
\end{equation}

\begin{equation}
    \label{eq:redEn}
    \frac{\partial \overline{T} }{\partial t}
    +
    \frac{\partial (\overline{u}_i \overline{T})}{\partial x_i}
    =
    \frac{\partial}{\partial x_j}
    \left(
      \alpha \frac{\partial \overline{T}}{\partial x_j} - q_j
    \right)
\end{equation}

\noindent where an overbar denotes a filtered variable, and $\delta_{ij}$ is
the Kronecker delta. The filtering operation is performed implicitly by the
mesh. The field variables $u$, $p$ and $T$ denote the velocity, pressure and
temperature, respectively, while $t$ denotes the  time. The constants are
$\rho_0$ -- the reference density, $T_0$ -- the reference temperature, $g$ --
the gravitational acceleration, $\beta$ -- the thermal expansion coefficient,
$\nu$ -- the kinematic viscosity and $\alpha$ -- the thermal diffusivity.
Finally, $\tau_{ij}$ and $q_j$ are the residual stress tensor and sub-grid heat
flux, respectively. 

The dynamic Smagorinsky eddy viscosity model provides turbulence closure
\cite{smagorinsky1963general,germano1991dynamic,lilly1992proposed}:

\begin{align}
    \tau_{ij}
    =
    -2c_s \Delta^2 \left| \overline{S} \right| \overline{S}_{ij}, & & q_j
    =
    \frac{(c_s \Delta^2 \left| \overline{S} \right|)}{Pr_t}
    \frac{\partial \overline{T}}{\partial x_j}.
\end{align}

\noindent where $\overline{S}_{ij}$ is the resolved strain-rate tensor, $\left|
\overline{S} \right| \equiv \sqrt{2\overline{S}_{ij}\overline{S}_{ij}}$,
$\Delta$ is the local filter width, $Pr_t$ is the turbulent Prandtl number, and $c_s$ is a model constant which is
allowed to vary spatially and temporally; i.e. $c_s = c_s(\mathbf{x},t)$. We
dynamically set $c_s$ according to the Germano-Lilly procedure
\cite{germano1991dynamic,lilly1992proposed}. This dynamic sub-grid model was
chosen since it is valid for relaminarised regions of the flow, as may be
encountered in stably stratified turbulence. 

In the solid domain, the transport of heat is accounted for via a reduced form
of Equation \ref{eq:redEn} in which the convective and sub-grid terms are zero.
The coupling between domains is achieved by enforcing consistency in both the
temperature and heat flux at the interface:

\begin{align}
    T_s 
    &
    =
    T_f \\
    \kappa_s \left.\frac{\partial \overline{T}}{\partial n}\right|_s
    &
    =
    -\kappa_f \left.\frac{\partial \overline{T}}{\partial n}\right|_f
\end{align}

\noindent where subscripts $(\cdot)_f$ and $(\cdot)_s$ denote the fluid and
solid domains, respectively, $\kappa$ is the thermal conductivity, and $n$ is the outward-pointing interface-normal.

The governing equations are discretised by the finite-volume method with
$\Delta$ taken as $V^{1/3}$ (where $V$ is the local cell volume), and are
solved with the CFD package Code\_Saturne (Version 5)
\cite{fournier2011optimizing}.

\subsection{Study Parameters}\label{sec:studyParam}

The fluid flow can be characterised by the Reynolds number ($Re$),
gravitational Richardson number ($Ri$) and Prandtl number ($Pr$):

\begin{align*}
  Re \equiv \frac{U D}{\nu},
  & &
  Ri \equiv \frac{g \beta (T_1 - T_0) D }{U^2},
  & &
  Pr \equiv \frac{\nu}{\kappa}, 
\end{align*}

\noindent where $U$ is the bulk velocity. In the present study, we set
$\mathit{Re}=10,000$, $\mathit{Ri}=2.23$, and $\mathit{Pr} = 6$. These parameters lead
to mixed convection flow conditions that are based on the study of
\cite{viollet1987observation}, and relevant to the nuclear industry (amongst
others). 

Note that by employing the Boussinesq approximation to account for buoyancy
(see Section \ref{sec:gov}), the centrifugal buoyancy is neglected. The impact
of this approximation can be quantified through consideration of the
centrifugal Richardson number:

\begin{align*}
  Ri_c \equiv \frac{\beta (T_1 - T_0) D}{R_c}.   
\end{align*} 

The ratio of $Ri_c$ to $Ri$ is then given by

\begin{align*}
  \frac{Ri_c}{Ri} = \frac{U^2}{g R_c}.
\end{align*} 

\noindent From the definition of $\mathit{Re}$, this can be rewritten as:

\begin{align*}
    \frac{Ri_c}{Ri} = \frac{ (\nu \mathit{Re} )^2}{D^2 g R_c},
\end{align*}

\noindent which for the present study is less than $9\times10^{-4}$ if  $D > 0.2 \mathrm{m}$. Similarly, the temperature-induced density difference is less than $0.5\%$ if $D > 0.2 \mathrm{m}$, and hence the Bousinessq approximation is reasonable for sufficiently large $D$.

The final dimensionless groups dictating the rate of conductive heat flow
between fluid and solid domains are the ratio of thermal diffusivities,
$\alpha_s / \alpha_f$, and the ratio of thermal conductivities, $\kappa_s / \kappa_f$. In the present study, we employ a ratio of $\alpha_s /
\alpha_f = 144.8$, and  $\kappa_s / \kappa_f = 123.5$, which is representative of water flowing within a steel
pipe. This is different to the experiments of \cite{viollet1987observation}, where 
altu-glass pipework was employed, and hence the thermal boundary condition was closer to adiabatic.
    
\subsection{Geometry and Mesh}
\begin{figure}[h]
  \centering
  \includegraphics[width=1.3\textwidth]{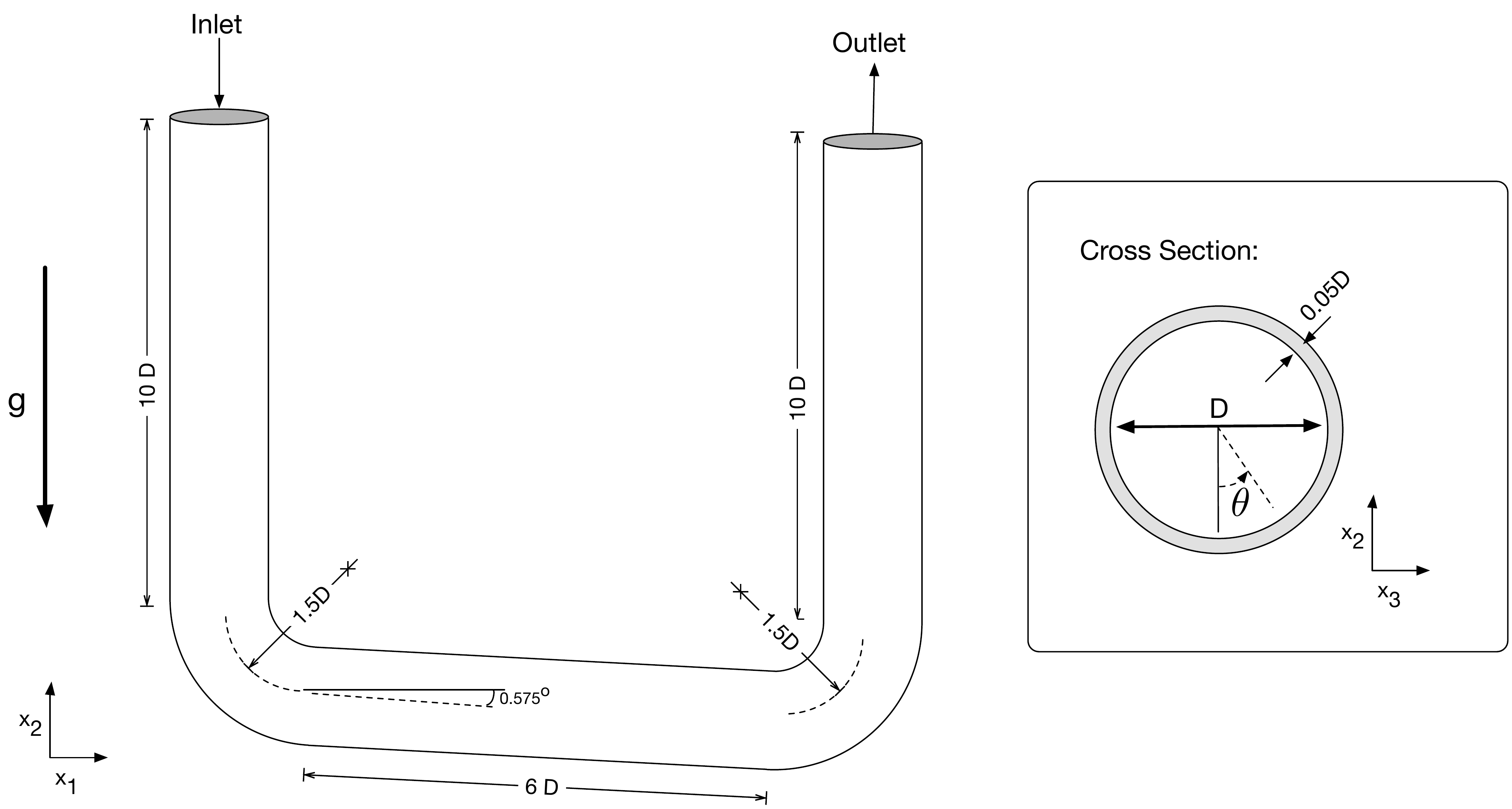}
  \caption{Schematic of the geometry}
  \label{fig:schematic}
\end{figure}

A schematic of the geometry is given in Figure~\ref{fig:schematic}. The
vertical inlet and outlet sections are $10 D$ in length, while the
near-horizontal section is $6 D$ in length with a $1\%$ downward slope. This slope is a feature of the Viollet case \cite{viollet1987observation}, and is an approximation of the pipework downstream of the steam-generator outlet of the Superph\'enix reactor. 

The radius of curvature, $R_c$, for both bends is $1.5 D$, while the wall-thickness is
$0.05 D$. All walls are smooth. This geometry is based on that of \cite{viollet1987observation}, but
has a longer vertical inlet section. The reason for this change is to allow
development of the inflow synthetic turbulence prior to the region of interest,
as will be discussed in Section \ref{sec:initialandboundary}.

The data presented herein has been computed on a block-structured mesh
comprising approximately $47M$ hexahedral cells ($43M$ and $4M$ for the fluid
and solid domains, respectively). The near-wall grid spacing was such that $y^+
< 0.2$ was maintained throughout the domain (with a corresponding $T^+ < 1.2$)
and hence no additional near-wall modelling or damping terms were required
(note the use of a dynamic sub-grid model, which precludes the need for
near-wall damping). 

A mesh sensitivity study has been conducted with a mesh comprising
approximately $24M$ cells. We observed no appreciable difference in low-order
statistics (mean velocity and temperature fields). The ensemble size (see Section \ref{sec:statistics} for details) for the
coarse mesh (ten runs) was insufficient for converged higher-order statistics,
hence the sensitivity of higher-order terms to the mesh cannot be ruled out.

\subsection{Initial and Boundary Conditions}
\label{sec:initialandboundary}
At the interface between fluid and solid domains, the no-slip condition
provides a boundary condition for the velocity, while a zero-gradient Neumann
condition is applied for the pressure. Unless otherwise stated, the interface temperature is set via the
conjugate transfer of heat between the domains, as described in Section
\ref{sec:gov}. At the external wall of the pipe, a zero gradient Neumann condition was applied for the temperature. Additional runs with an adiabatic wall thermal boundary condition have also been performed for comparison purposes.

Correlated inflow data, with prescribed first- and second-order statistics was
generated via the Synthetic Eddy Method (SEM)
\cite{jarrin2006synthetic,skillen2016accuracy}. The statistical input to the
SEM was generated via a precursor Reynolds Averaged Navier-Stokes (RANS)
simulation of fully-developed pipe-flow, also at $Re=10,000$, in which the Elliptic Blending Reynolds Stress Model
(EBRSM) \cite{manceau2002elliptic} provided turbulence closure. Note that
we deliberately do not account for the change in the mean velocity profile and
Reynolds stress tensor at the inlet due to upstream thermal stratification. Our
aim is to assist reproducibility, and hence a simple stationary condition for
the velocity and Reynolds stresses was favoured. Moreover, tests showed that
the synthetic turbulence reached a mature state within $\sim 2 D$ (assessed via
the development of wall skin-friction coefficient) -- well before the first
bend. This development length is in line with previous work at a similar
Reynolds number \cite{skillen2016accuracy}.

The flow was initialised by conducting an isothermal LES computation at
reference temperature $T_0$. The solution was time-marched to a
statistically-steady state, to provide initial conditions to the main
computation. At time $\tilde{t} (\equiv U t / D) = 0$, we initiated the thermal
transient by linearly increasing the inlet temperature to a value $T_1$. This
linear ramp acts until $\tilde{t} = 7.5$. For all times $\tilde{t} > 7.5$, the
inlet temperature remains fixed at $T_1$. The magnitude of the temperature
difference, $(T_1 - T_0)$, is dictated by the Richardson number, $Ri$.  

At the outlet, zero-gradient Neumann conditions were applied for all variables,
with the exception of the pressure which had a Dirichlet condition applied.

\subsection{Statistical Data} \label{sec:statistics}
Due to the transient nature of the flow, averaging is conducted via an
ensemble. We generate a total of forty realisations of the flow, each by
seeding the random number generator of the synthetic inflow method differently.
This was done at the start of the isothermal initialisation computation. This
ensemble size is sufficient for well-converged low-order statistics, and
reasonable convergence of higher-order statistics (enough to be useful in the
assessment of unsteady-RANS models, for example). The convergence of the
ensemble is assessed via Table~\ref{tab:convergence}.

\section{Results and Discussion}\label{sec:results}
\subsection{Flow and Heat Transfer Overview}
\begin{figure}
  \centering
  \begin{subfigure}{0.48\textwidth}
    $\tilde{t}=0$
    \centering
    \includegraphics[width=1.3\textwidth]{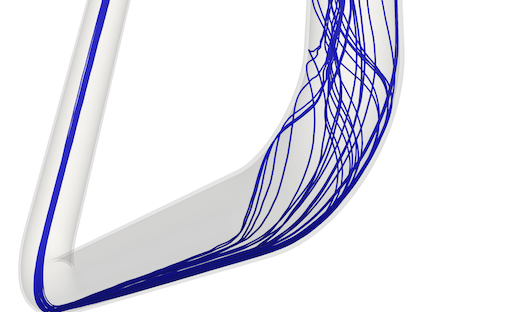}
  \end{subfigure}
  ~
  \begin{subfigure}{0.48\textwidth}
    $\tilde{t}=15$
    \centering
    \includegraphics[width=1.3\textwidth]{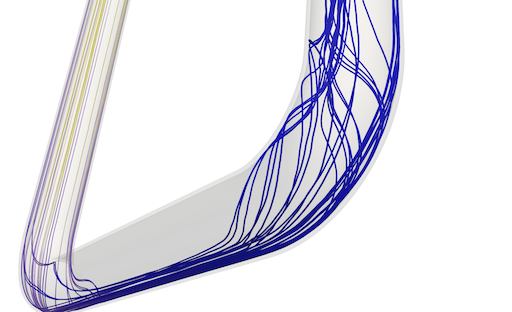}
  \end{subfigure}

  \begin{subfigure}{0.48\textwidth}
    $\tilde{t}=30$
    \centering
    \includegraphics[width=1.3\textwidth]{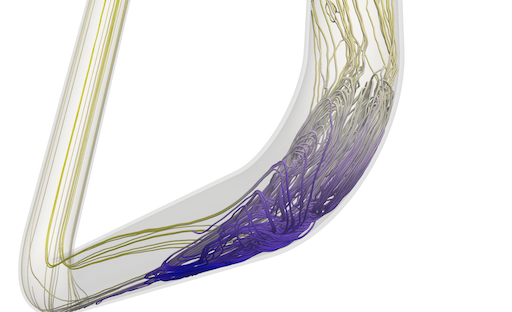}
  \end{subfigure}
  ~
  \begin{subfigure}{0.48\textwidth}
    $\tilde{t}=45$
    \centering
    \includegraphics[width=1.3\textwidth]{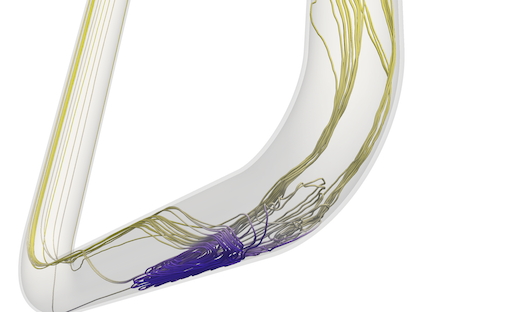}
  \end{subfigure}

  \caption{(Colour online) Ensemble averaged flow streamlines, coloured by
  temperature. All time-frames have their streamlines seeded at the same
  location. $T_0$: Blue. $T_1$: Yellow.}
  \label{fig:sl}
\end{figure}

In Figure~\ref{fig:sl}, ensemble averaged streamlines showing the temporal
evolution of the flow are presented. The formation of a large buoyancy-induced
recirculation region is observed, qualitatively matching the observations of
\cite{viollet1987observation}. It can be seen that the thermal stratification
persists for several dimensionless time-units, indicating limited heat transfer
into the recirculation region.

\begin{figure}
  \centering
  \includegraphics[width=1.5\textwidth]{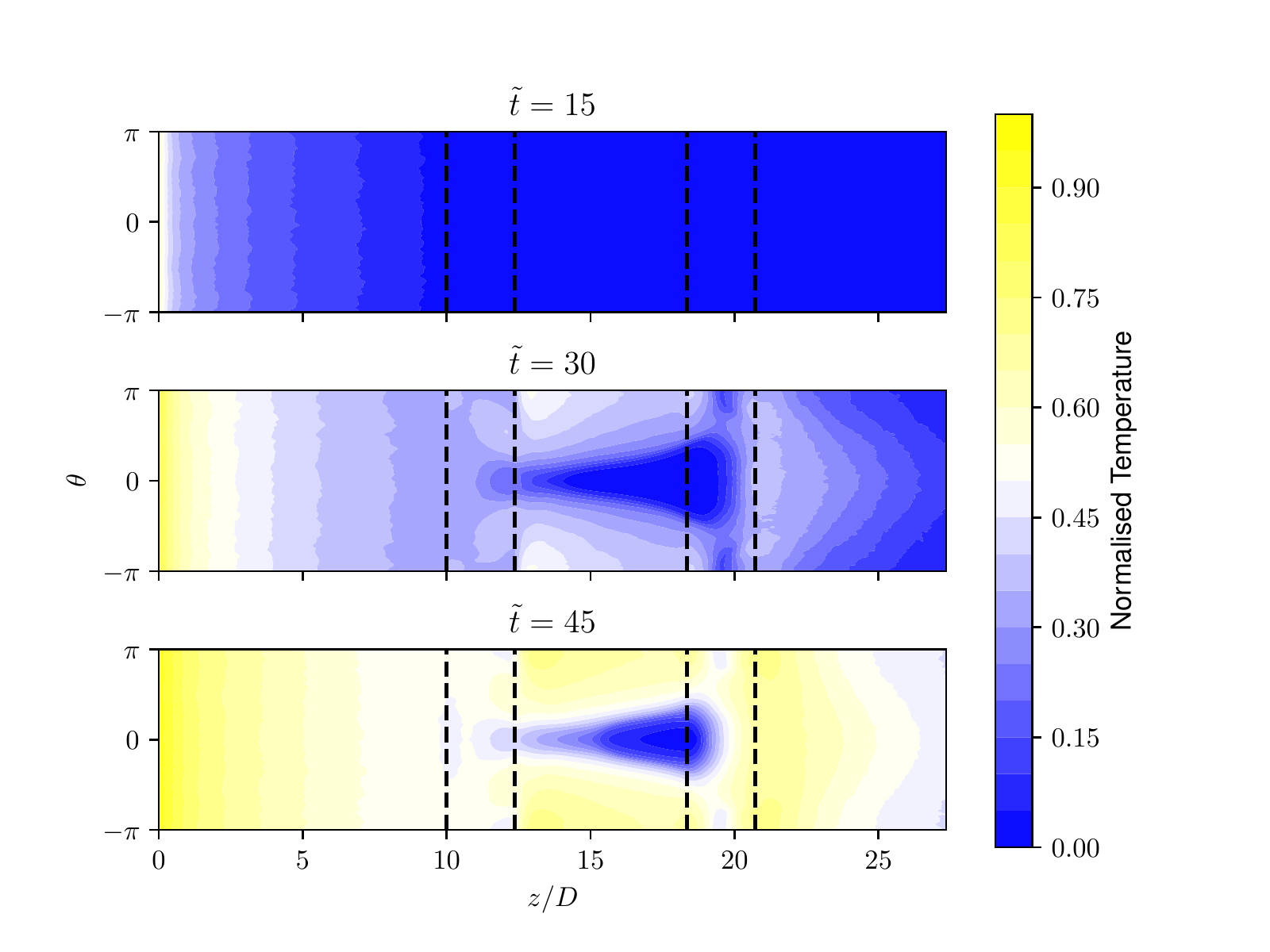}
  \caption{(Colour online) Flattened interface temperature contours
  (cylindrical coordinate system). Dashed lines denote locations of the two
  bends. The inlet is at $z/D=0$.}
  \label{fig:flatten_T}
\end{figure}

Figure~\ref{fig:flatten_T} shows contours of the fluid-solid interface
temperature evolution. At times $\tilde{t}=30$ and $\tilde{t}=45$, the dominant
feature of the plots is the stratification-induced cold wall region along the
bottom of the near-horizontal section. Also noteworthy is the temperature lag
of the wall. This is caused by the wall's significant thermal inertia. For
instance, by $\tilde{t}=15$ the thermal transient has propagated beyond the
first bend (the thermal-front is located within the near-horizontal section).  Yet the interface temperature at $z/D \gtrsim 9$ is still at (or very close to)
$T_0$. For the vertical pipe section, at $z/D < 10$ the interface temperature
is significantly below $T_1$ even by $\tilde{t}=30$, and has still not reached
the final temperature by $\tilde{t}=45$. Clearly this interface temperature
is not correctly captured with an adiabatic wall boundary condition
(where the thermal inertia is zero). The role this thermal inertia has upon the
flow-physics will become apparent in the following discussion.

\begin{figure}
  \centering
  \includegraphics[width=1.5\textwidth]{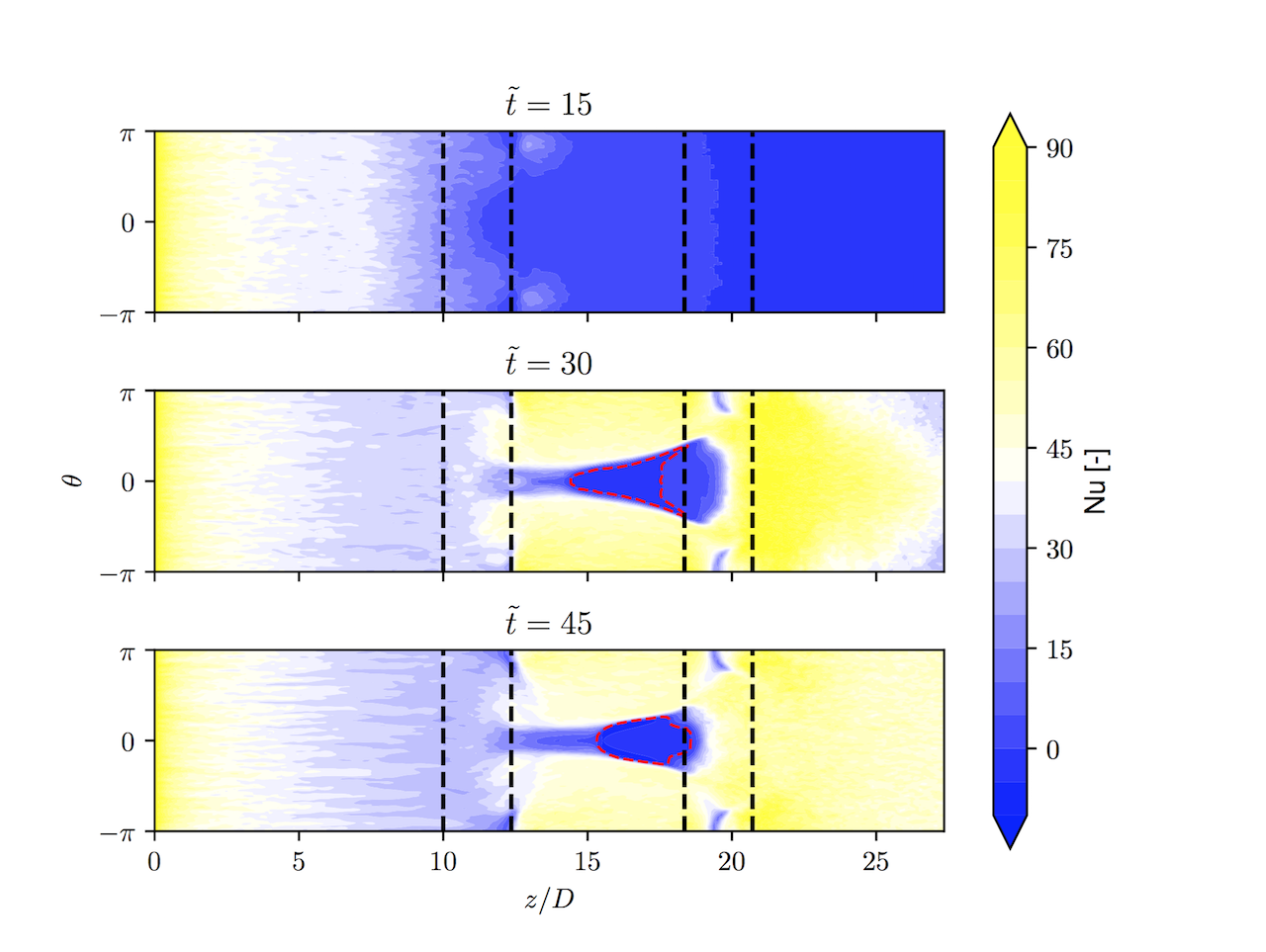}
  \caption{(Colour online) Flattened interface Nusselt number contours
  (cylindrical coordinate system). Zero contour highlighted by red-dashed
  line. Black dashed lines denote locations of the two bends. The inlet is at
  $z/D=0$.}
  \label{fig:flatten_nu}
\end{figure}

In Figure~\ref{fig:flatten_nu}, we present contours of the fluid-solid
interface Nusselt number evolution. Interestingly, negative Nusselt number
regions are observed towards the end of the transient. This indicates that one mode of heat
transfer into the cold recirculation region is via the solid domain; as the
upper portion of the near-horizontal pipe wall heats up, heat is transferred
circumferentially through the solid domain by conduction. Below the cold
recirculation region, some of this heat is subsequently transferred back to the
fluid from below, leading to a negative Nusselt number.

\begin{figure}
  \centering
  \includegraphics[width=1.3\textwidth]{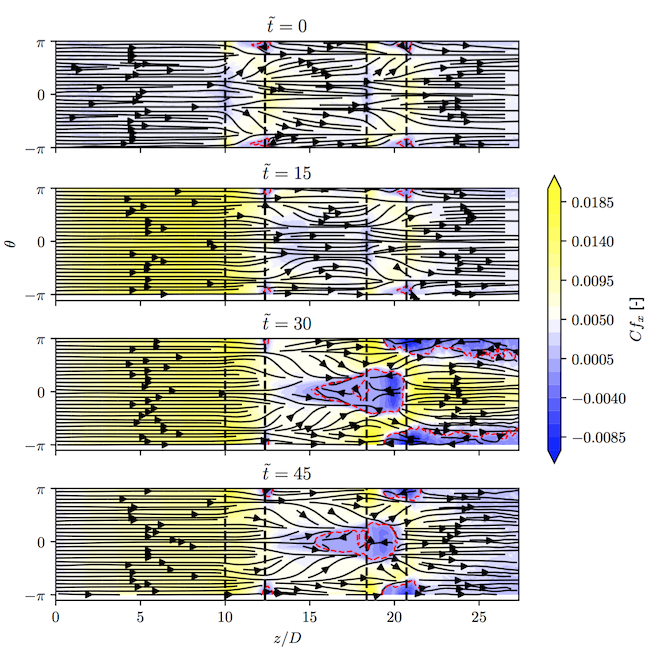}
  \caption{(Colour online) Interface skin-friction lines coloured by streamwise
  skin-friction component. The zero-contours of streamwise skin-friction are
  highlighted by the red-dashed lines. Dashed black lines denote locations of the
  two bends. The inlet is at $z/D=0$.}
  \label{fig:flatten_wss}
\end{figure}

Wall shear-stress vectors are shown in Figure~\ref{fig:flatten_wss}. At all
times, a small recirculation region is observed on the inner radius of each
bend (apparent by negative streamwise skin-friction). It is interesting to note
that for time $\tilde{t}=30$, the reversed flow after the second bend continues
all the way to the exit of the domain (note the negative streamwise
skin-friction at $\theta \approx \pi$, and $z>20$). This is due to rapid flow
acceleration as the thermal transient reaches the downstream vertical section
and experiences unstable stratification. This generates a buoyant plume, and
rapid flow acceleration. Fluid entrainment by this plume leads to a
recirculation region, in accordance with mass conservation.  This is in
qualitative agreement with the observations of \cite{viollet1987observation}.  

\begin{figure}
  \begin{minipage}{0.78\textwidth}
    \centering
    \begin{subfigure}{0.31\textwidth}
      $\tilde{t}=0$
      \centering
      \includegraphics[width=\textwidth]{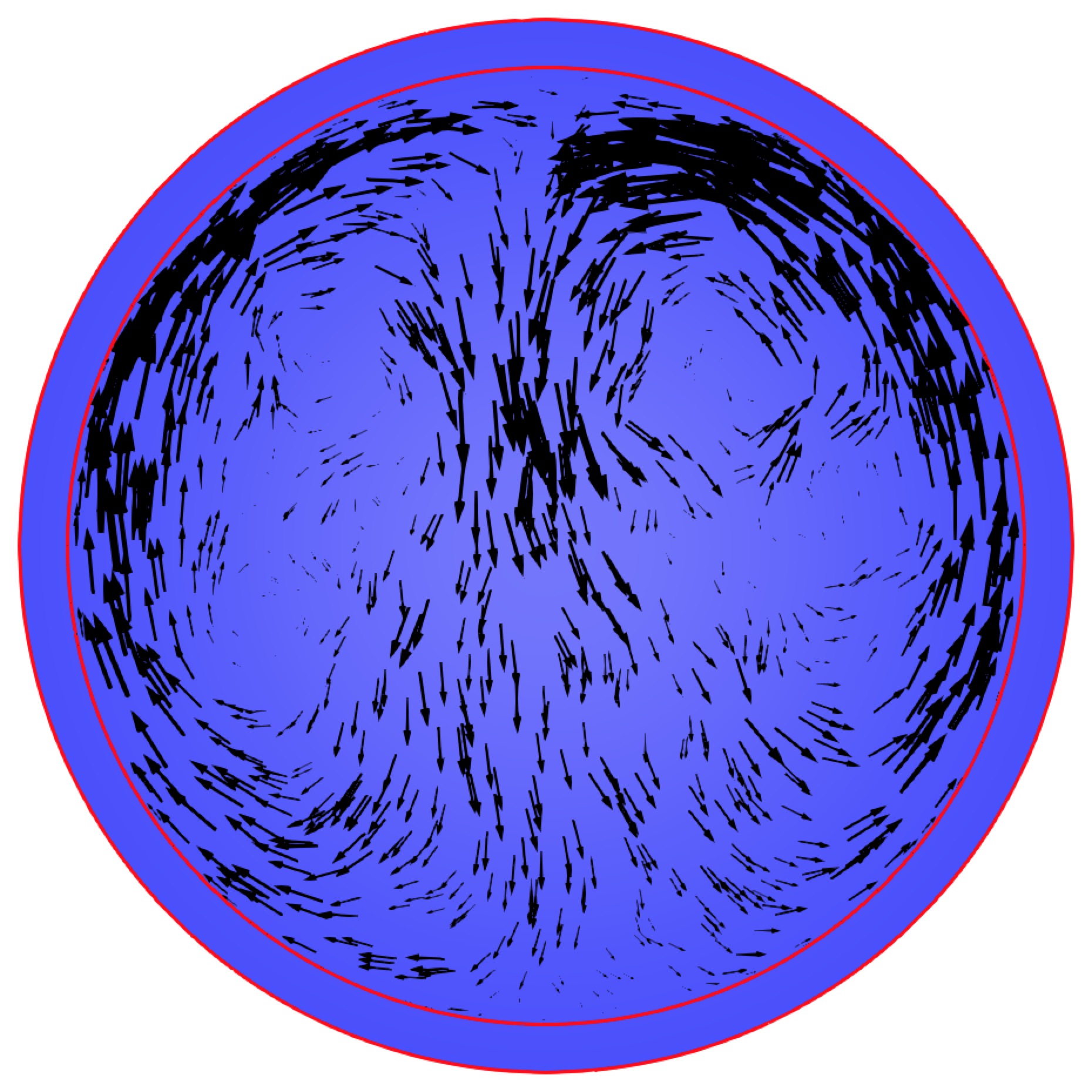}
    \end{subfigure}
    ~
    \begin{subfigure}{0.31\textwidth}
      $\tilde{t}=15$
      \centering
      \includegraphics[width=\textwidth]{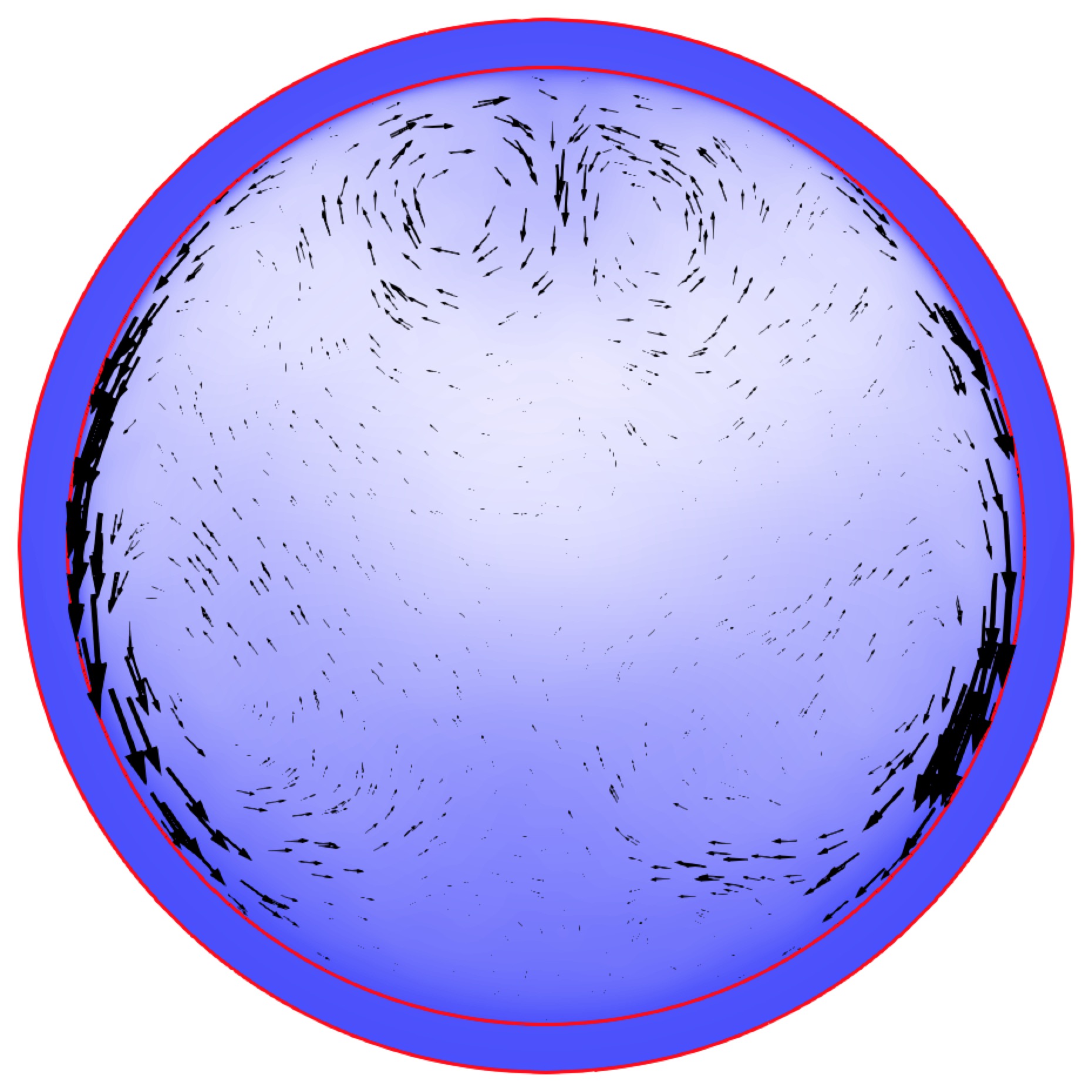}
    \end{subfigure}
    ~
    \begin{subfigure}{0.31\textwidth}
      $\tilde{t}=30$
      \centering
      \includegraphics[width=\textwidth]{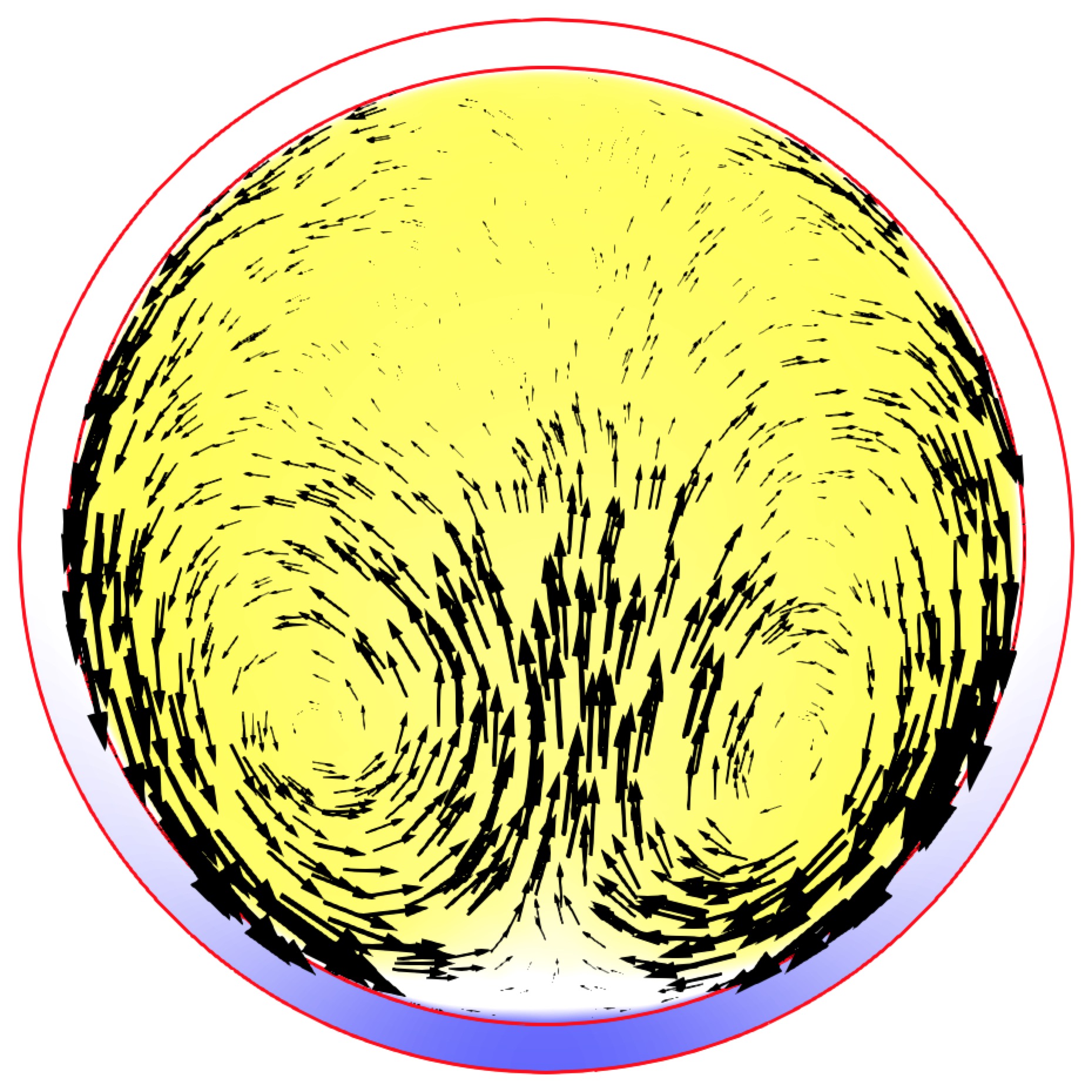}
    \end{subfigure}
    
    \begin{subfigure}{0.31\textwidth}
      $\tilde{t}=45$
      \centering
      \includegraphics[width=\textwidth]{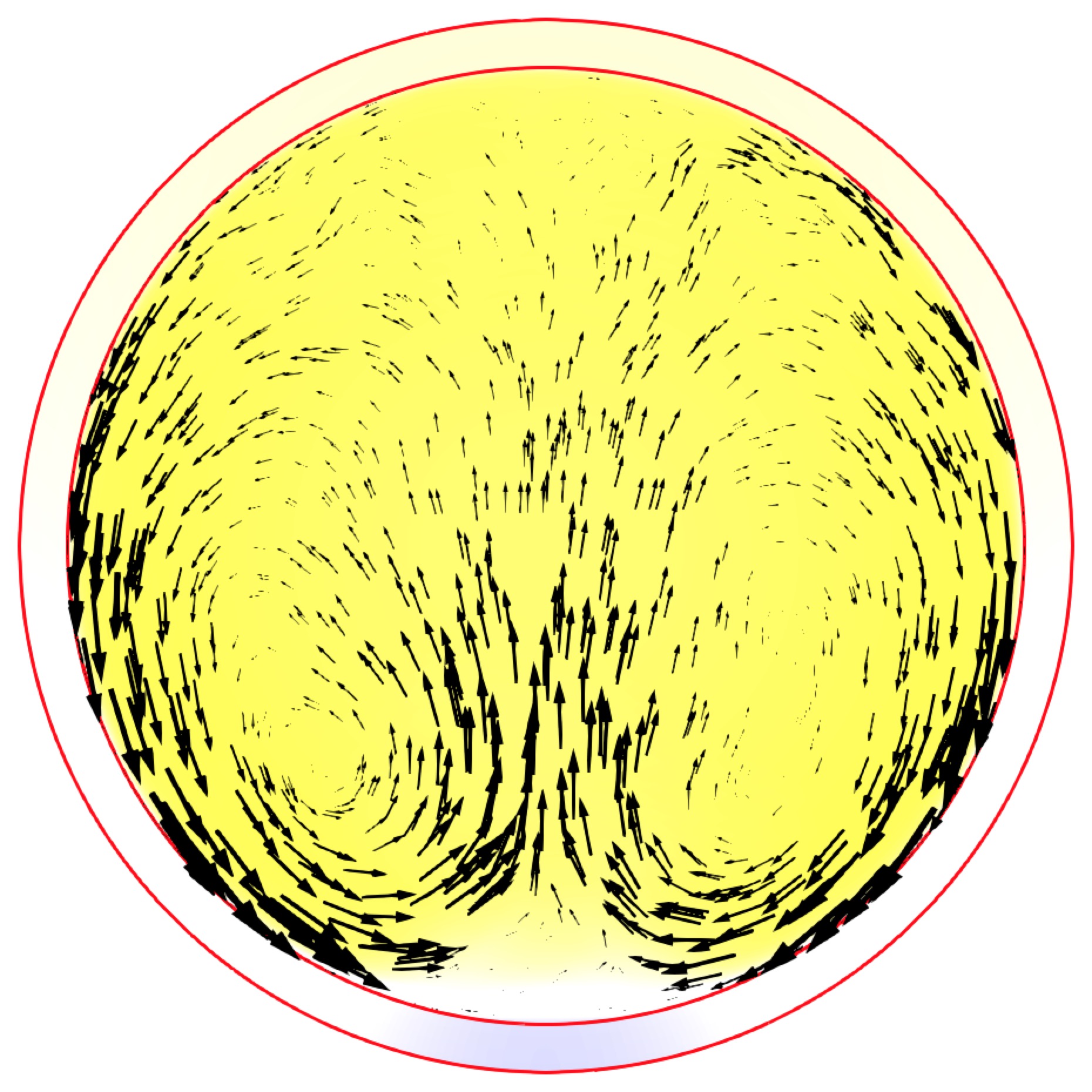}
    \end{subfigure}
    ~
    \begin{subfigure}{0.31\textwidth}
      $\tilde{t}=70$
      \centering
      \includegraphics[width=\textwidth]{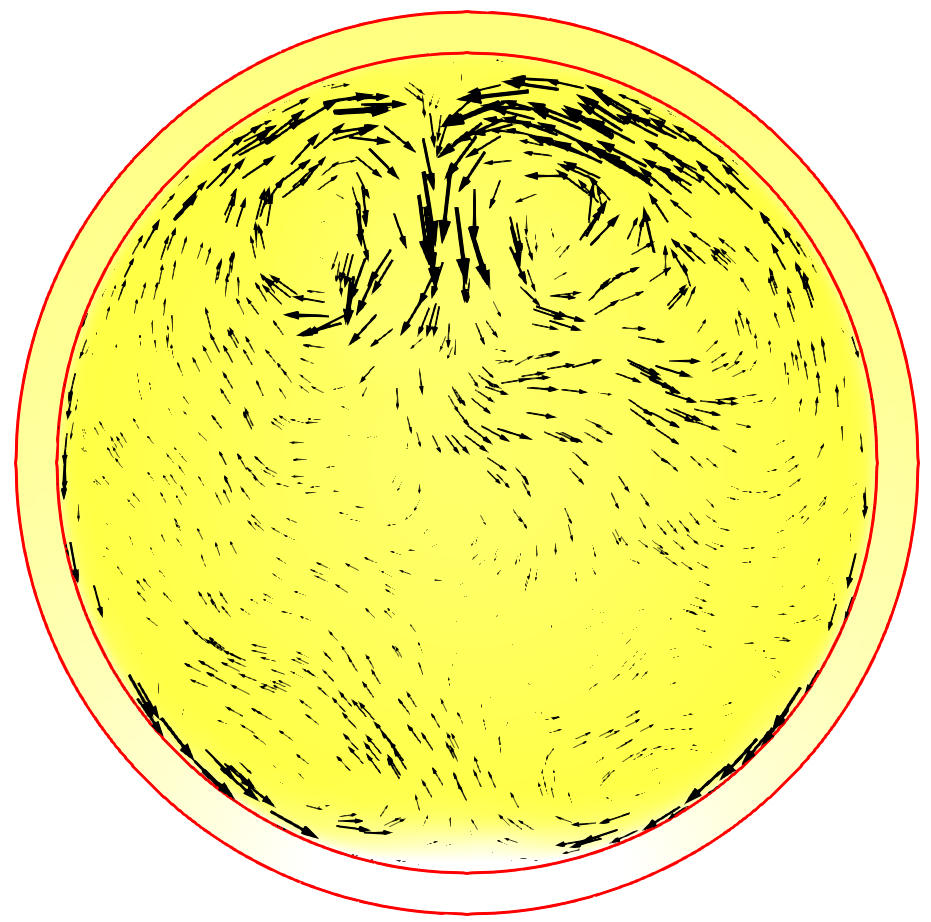}
    \end{subfigure}

  \end{minipage}
  \begin{minipage}{0.2\textwidth}   
    \includegraphics[width=\textwidth]{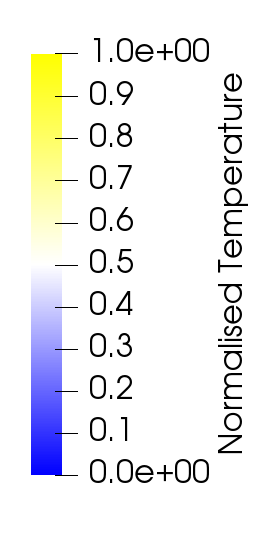}
  \end{minipage}

  \caption{(Colour online) Ensemble averaged secondary flow, $1 D$ downstream
  of the first bend, coloured by normalised temperature. Red outline denotes the boundaries of the solid domain.}
  \label{fig:secondary_flow}
\end{figure}

It can be seen from Figure~\ref{fig:flatten_wss}, at time $\tilde{t}=0$ there
is a secondary transport (by the Dean vortices) from $\theta=0$ towards
$\theta=\pi$ after each bend. The centripetal force associated with the pipe
bend causes low-inertia near-wall fluid to be transported towards the centre of
curvature (at $\theta = \pi$). Interestingly, by $\tilde{t}=15$, the secondary
transport is in the opposite direction to that which would be expected from
Dean vortices. To investigate this further, we plot vectors of the secondary
flow at a section $1 D$ downstream of the end of the first bend (see
Figure~\ref{fig:secondary_flow}). At time $\tilde{t}=0$, classical Dean
vortices are observed. At $\tilde{t}=15$ the secondary flow is suppressed, and
starting to reverse. By $\tilde{t}=30$, a strong counter-rotating vortex pair
is observed, with opposite sign to that at time $\tilde{t}=0$. This can be
explained by consideration of the thermal inertia of the wall, and the
buoyancy-driven secondary flow this induces. As the hot-front enters into the
near-horizontal section of the pipe, the fluid at the core of the pipe is
hotter and lighter than the cold near-wall fluid (cold, due to the large
thermal-inertia of the wall). Buoyancy-driven density currents are initiated,
which dominate the centripetal force. The cold near-wall fluid  sinks,
generating a pair of counter-rotating vortices with opposite sign to that of
the Dean vortices. This is qualitatively similar to the results of Kurnia et
al. \cite{kurnia2016numerical}, with the exception that their density gradient
was induced by fixing the wall temperature, while ours is due to the
combination of a thermal transient and the wall's thermal inertia. 

By $\tilde{t}\approx 70$, we observe a second reversal in the secondary flow as the wall 
heats up, and the centripetal force again dominates. This is apparent in the final frame of Figure \ref{fig:secondary_flow}, where two pairs of counter-rotating vortices can be seen; one due to 
the centripetal force as the Dean vortices re-establish, and the other (smaller) pair due to buoyancy induced by the thermal inertia of the wall. Eventually, we would expect 
only Dean vortices, identical to time $\tilde{t}=0$, as the flow becomes isothermal again at temperature $T_1$. The computations were not run to the new isothermal state. 

\begin{figure}
    \begin{subfigure}{0.41\textwidth}
      Conjugate heat transfer
      \centering
      \includegraphics[width=\textwidth]{sec_30_crop}
    \end{subfigure}
    ~
    \begin{subfigure}{0.41\textwidth}
      Adiabatic
      \centering
      \includegraphics[width=\textwidth]{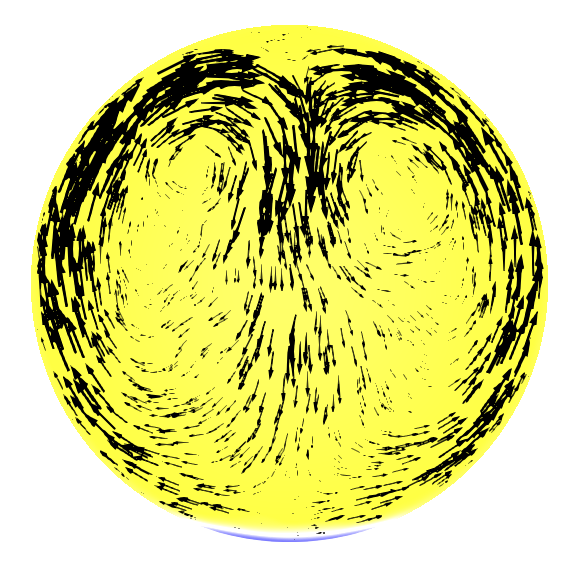}
    \end{subfigure}
    \caption{(Colour online) Ensemble averaged secondary flow at $\tilde{t}=30$, $1 D$ downstream of the first bend, coloured by normalised temperature (colour bar as in Fig. \ref{fig:secondary_flow}). Red outline denotes the boundaries of the solid domain.}
  \label{fig:adcomp}
\end{figure}

To confirm that the reversal of the secondary flow is indeed due to the thermal inertia of the wall, we have also conducted an ensemble of adiabatic runs. All details of the adiabatic computations are identical to those of the conjugate heat transfer runs (mesh, ensemble size, etc.) with the exception of the thermal boundary condition on the inner wall. In Figure~\ref{fig:adcomp}, the secondary flow at time $\tilde{t}=30$ is compared for the two different thermal boundary conditions. It can be seen that in the adiabatic case, the secondary flow is in the opposite direction to that of the conjugate heat transfer case, thus confirming the thermal inertia of the wall does indeed reverse the sign of the secondary vortices. This highlights the importance of including the solid domain for transient computations of mixed or natural convection, where the thermal inertia of the wall is high .

\begin{figure}
  \includegraphics[width=\textwidth]{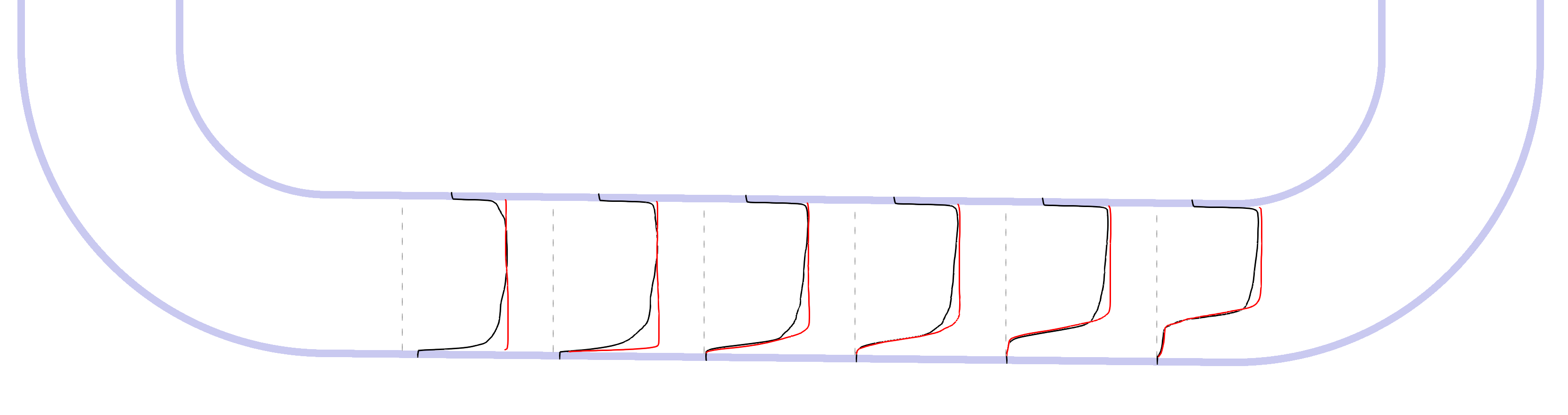}
    \caption{(Colour online) Profiles of the ensemble averaged temperature on the symmetry plane, at $\tilde{t}=30$ (Conjugate heat transfer -- Black lines. Adiabatic boundary condition -- red lines.). The profiles are taken 
 at $x/D=2$ to $7$ in increments of $1$, with the locations denoted by the dashed lines lines.}
  \label{fig:profiles}
\end{figure}

Profiles of the ensemble-averaged temperature on the symmetry plane,  at $\tilde{t}=30$,  
are shown in Figure~\ref{fig:profiles}. From this figure, the extent of the stratification region is apparent. The impact  an adiabatic boundary condition has is also apparent from this figure. Significant differences in the thermal field are apparent between the conjugate and adiabatic cases, highlighting the major influence this secondary flow has upon the global heat transfer characteristics of the pipework. Again, this highlights the importance of including the solid domain. The remainder of this paper focuses on the conjugate heat transfer results.

\subsection{Vorticity Budgets}
\begin{figure}
  \centering
  \includegraphics[width=1.5\textwidth]{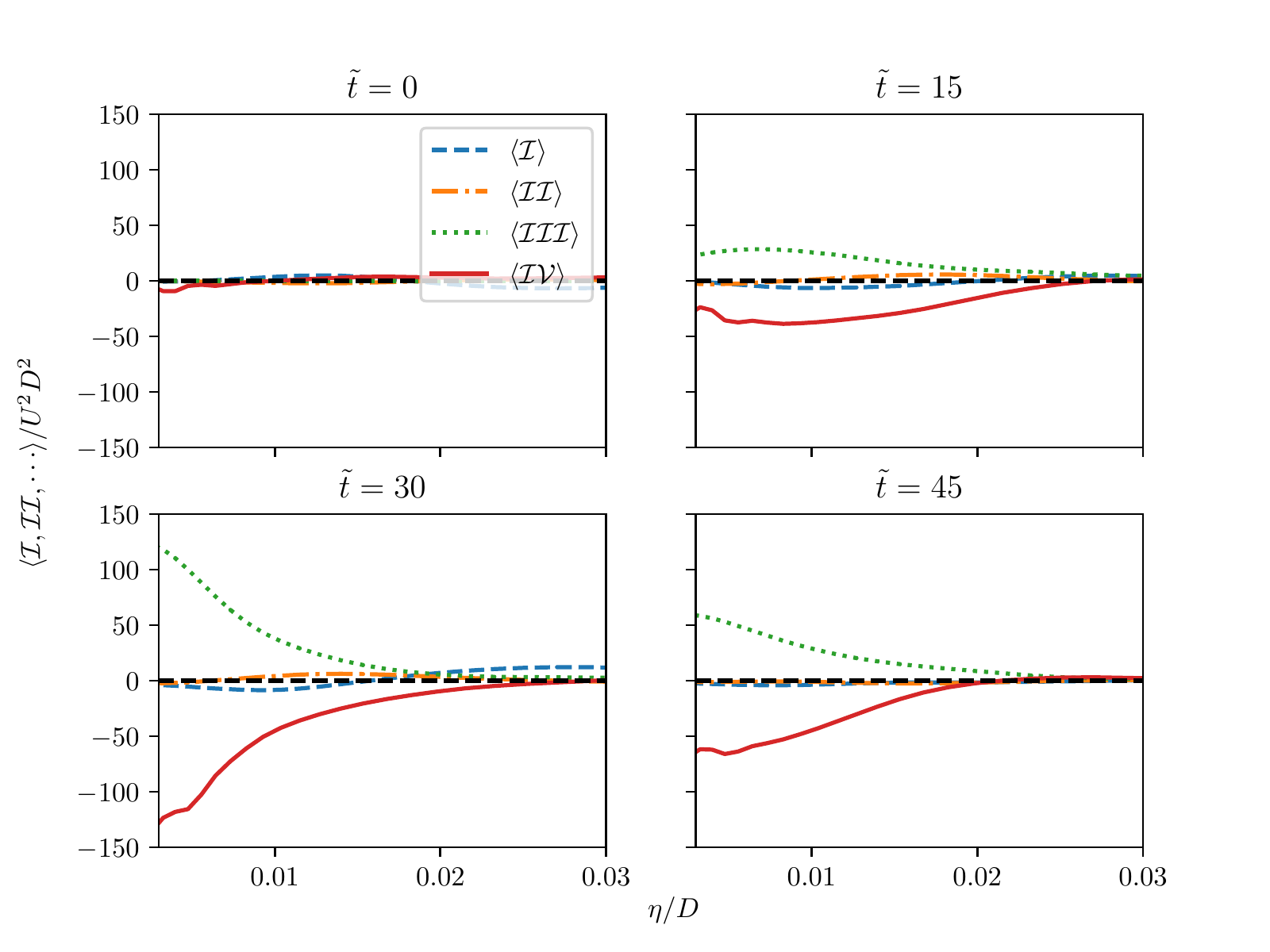}
  \caption{(Colour online) Ensemble averaged streamwise vorticity budgets, $1
  D$ downstream of the first bend. Profiles are in the wall-normal direction
  ($\eta$), at $\theta=\pi/2$.}
  \label{fig:vort_budgets}
\end{figure}

We study the budgets of the vorticity transport equation in order to elucidate
further the mechanism of secondary-flow reversal. The vorticity transport
equation can be obtained by applying the curl operator to the momentum
transport equation, and is written as

\begin{align}
  \underbrace{\frac{D \omega_i}{D t}}_{\mathcal{I}} 
  =
  \underbrace{\omega_j \frac{\partial u_i}{\partial x_j}}_{\mathcal{II}}
  +
  \underbrace{
    \epsilon_{ijk}\frac{1}{\rho_0}
    \frac{\partial (\beta T)}{\partial x_j}
    \frac{\partial p}{\partial x_k}
  }_{\mathcal{III}}
  +
  \underbrace{
    \nu\frac{\partial^2 \omega_i}{\partial x_j \partial x_j}
  }_{\mathcal{IV}},
\label{eq:omega}
\end{align}

\noindent where $\epsilon_{ijk}$ is the Levi-Civita symbol. On the right hand
side, $\mathcal{II}$ is the vortex stretching term, $\mathcal{III}$ is the
baroclinic torque, and $\mathcal{IV}$ is the viscous diffusion of vorticity.
The baroclinic term is active when the density gradient is misaligned with the
pressure gradient. In our case, at the exit of the first bend, the pressure
gradient balances the centrifugal and hydrostatic forces, and hence has a large
vertical component. Meanwhile, the density gradient acts in approximately the
wall-normal direction, due to the thermal inertia of the wall. At
$\theta=\pi/2$, we would thus expect the baroclinic term to be large as the
transient passes, but before the wall has fully heated. In
Figure~\ref{fig:vort_budgets}, we plot the vorticity budgets in the wall-normal
direction at $\theta=\pi/2$. It can be seen that the baroclinic term,
$\mathcal{III}$, is the dominant source of vorticity production as the thermal
transient passes. This vorticity is redistributed by viscous diffusion
($\mathcal{IV}$) which balances the baroclinic production term. Note that the
residual of Equation (\ref{eq:omega}) is not zero, as is apparent from studying
Figure~\ref{fig:vort_budgets}. This is due to the lack of convergence of the
higher-order terms (the convective term in $\mathcal{I}$, and the vortex
stretching term, $\mathcal{II}$) due to the sample size available for ensemble
averaging. Despite this, it is apparent that the baroclinic term (which is well
converged) is the dominant mode of vorticity production for times where the
thermal transient is present. 

\subsection{Swirl-Switching}
\begin{figure}
  \includegraphics[width=1.5\textwidth]{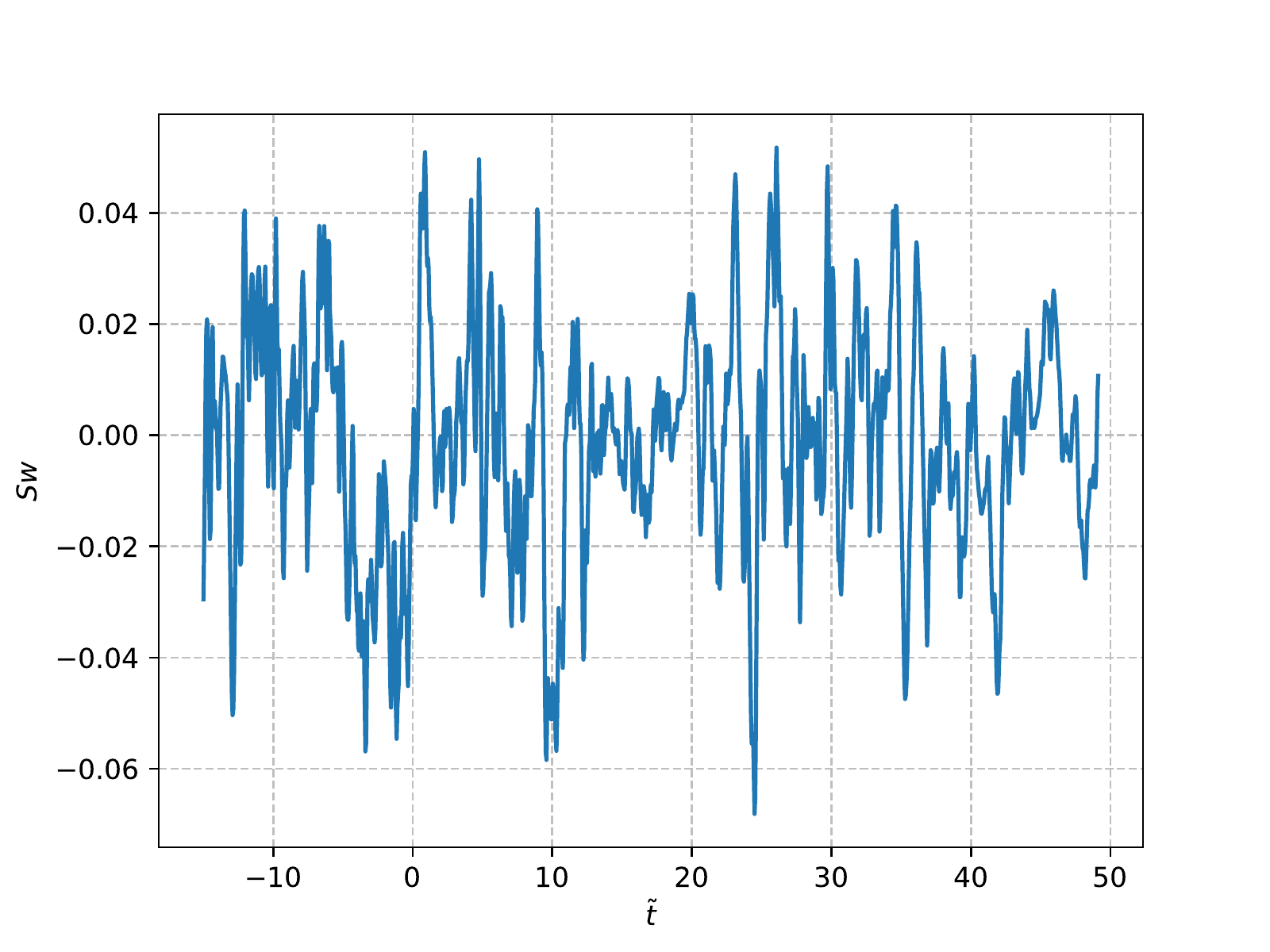}
  \caption{Evolution of the swirl-number, $Sw$, with time. Swirl number is
  integrated at the cross-section $1 D$ downstream of the first elbow. Negative 
values of $\tilde{t}$ correspond to the isothermal flow, prior to the transient}
  \label{fig:sw_time}
\end{figure}

Swirl-switching can affect the fatigue life of pipes where the flow is
subjected to a bend. We are interested in assessing how the thermal transient
affects this phenomenon. Swirl-switching can be characterised through the
swirl number, $Sw$, which is defined as the ratio of the flux of angular
momentum in the axial direction and the flux of axial momentum in the axial
direction, normalised by pipe radius:

\begin{align}
  Sw = \frac{
    \int r \rho U_\theta U_{ax} dA
  }
  {
    D/2 \int \rho \left| U_{ax} \right| U_{ax} dA
  },
\end{align}
\noindent where $U_\theta$ and $U_{ax}$ are the tangential and axial velocity
components, respectively. Figure~\ref{fig:sw_time} shows the temporal evolution of $Sw$
integrated over a plane $1 D$ downstream of the first bend. It is apparent from
the plot that the amplitude of the swirl number decreases between $\tilde{t}
\approx 12 - 20$. This is when the thermal transient reaches the
plane under consideration, and the secondary-flow starts to change direction.

\begin{figure}
  \includegraphics[width=1.3\textwidth]{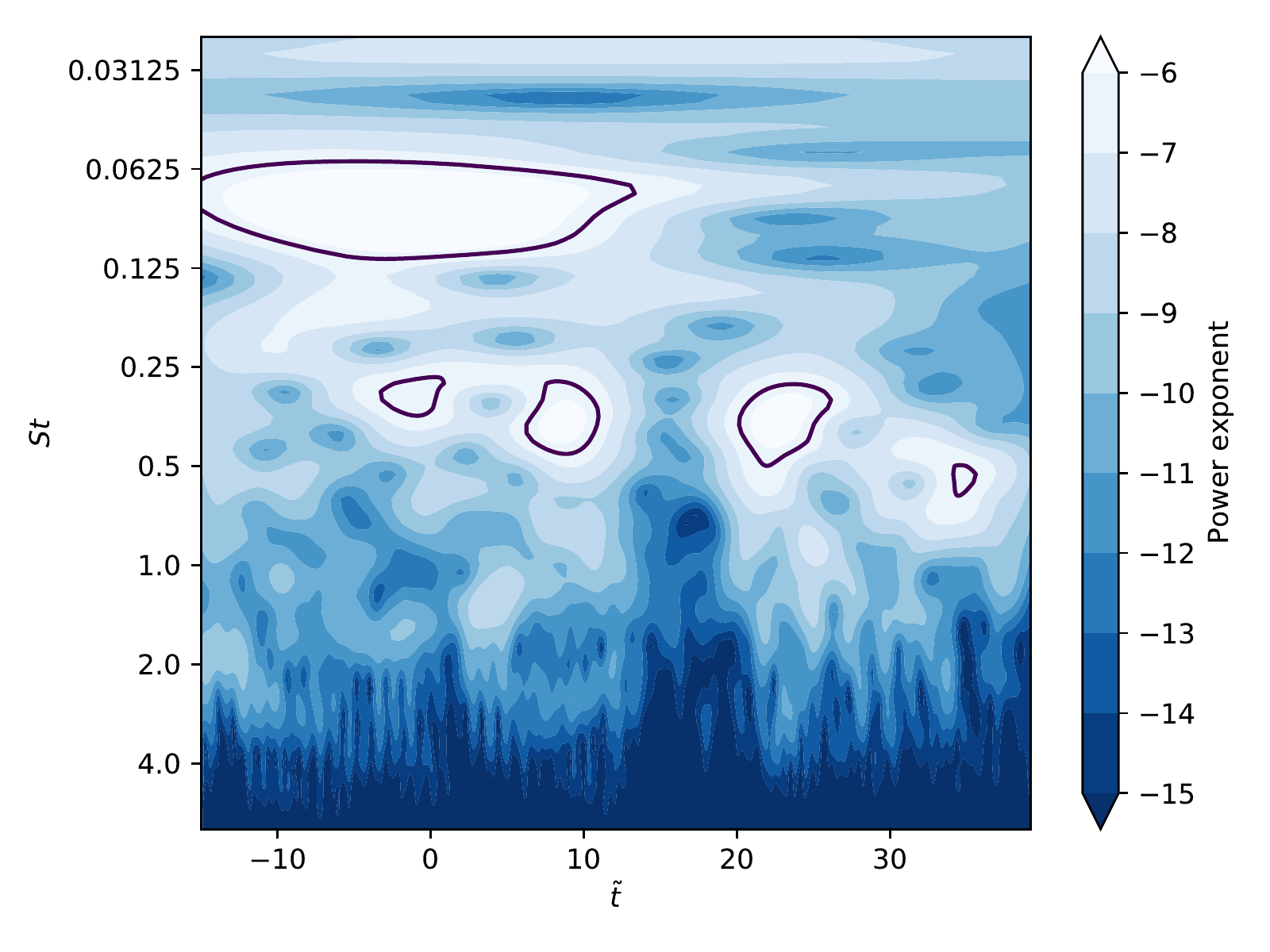}
  \caption{(Color online) Continuous wavelet transform scalogram showing the
  evolution of Strouhal number, ($St$), based upon the swirl number, ($Sw$). Swirl number is integrated at
  the cross-section $1 D$ downstream of the first elbow. Black contour lines show
  statistically significant spectral peaks, with a $95\%$ significance level.}
  \label{fig:cwt}
\end{figure}

In order to assess the frequency of the swirl-switching, we perform a
continuous wavelet transform (CWT) to the $Sw(\tilde{t})$ signal, using a
Morlet mother wavelet \cite{mallat1999wavelet}. This allows us to transform the time signal to the frequency domain, and is applicable to a non-stationary signal. To reduce the noise in the spectral power estimation, we
ensemble average the CWT coefficients over an ensemble size of three runs (note
the smaller ensemble size than for the flow statistics, due to the availability
of data). This process is analogous to Welch's method for spectral density
estimation for a stationary signal. The resulting scalogram is presented in
Figure~\ref{fig:cwt}. It can be seen from the figure that prior to the arrival
of the thermal transient ($\tilde{t} \lesssim10$) there are two dominant
frequencies at $St\equiv f D / U \approx0.1$ and $St\approx 0.3$. This is in
agreement with prior studies of the flow in pipe bends
\cite{hellstrom2013turbulent}.  As the secondary flow reversal progresses ($12
\lesssim \tilde{t} \lesssim 20$), the spectral power is significantly reduced
over a broadband range of frequencies. Indeed, there are no statistically
significant peaks in the spectral power estimation. As the reversed state of
secondary flow establishes itself (at $\tilde{t} \gtrsim 20$), there appears to
be only one dominant frequency at $St\approx 0.3$. The low frequency mode that
was present for $\tilde{t} \lesssim 10$ has been suppressed. Since this low
frequency mode is thought to originate from the large-scale structures upstream
of the bend \cite{wang2018direct}, one possible explanation for this
observation is the turbulence-suppressing effects of stable stratification,
which reduce the kinetic energy contained within the large structures.

\begin{figure}
  \includegraphics[width=\textwidth]{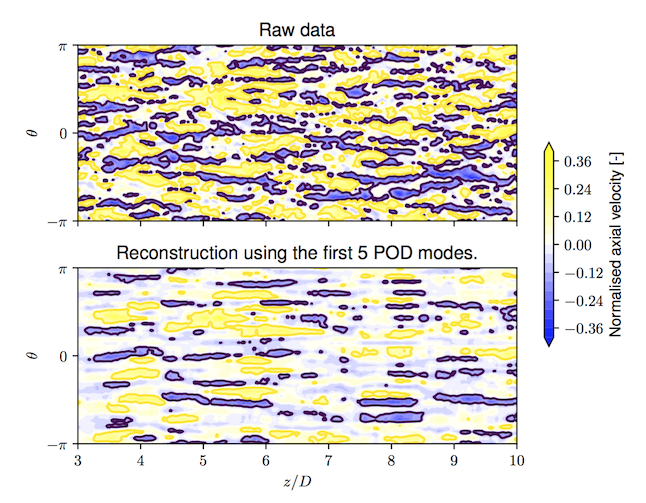}
\caption{(Color online) Top: Contours of $U_{ax}/U$ at time $\tilde{t}=0$, taken $0.05D$ from the pipe wall. Bottom: A reconstruction using the first five POD modes, highlighting the most energetic structures. In both plots, the $\pm0.1$ contours are emphasised by darkened contour lines.}
  \label{fig:structures_0}
\end{figure}

\begin{figure}
  \includegraphics[width=\textwidth]{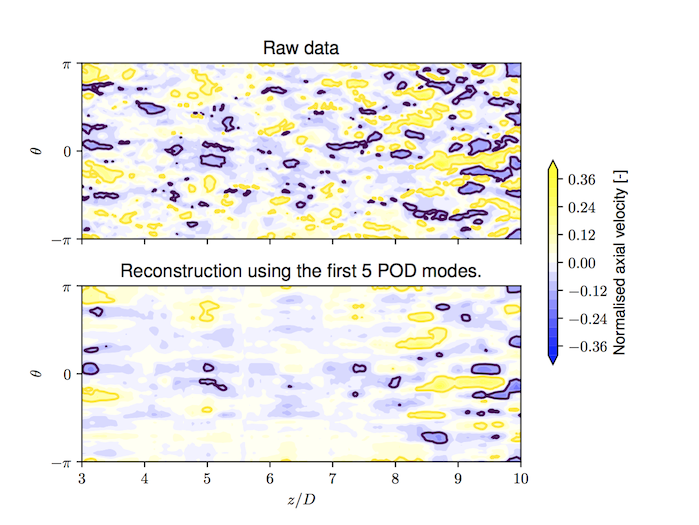}
\caption{(Color online) Top: Contours of $U_{ax}/U$ at time $\tilde{t}=30$, taken $0.05D$ from the pipe wall. Bottom: A reconstruction using the first five POD modes, highlighting the most energetic structures. In both plots, the $\pm0.1$ contours are emphasised.}
  \label{fig:structures_30}
\end{figure}

Figures \ref{fig:structures_0} and \ref{fig:structures_30} show contours of the 
normalised axial velocity $0.05D$ from the wall, for times $\tilde{t}=0$ and $\tilde{t}=30$, respectively.
Also presented in the plots is a reconstruction using the first five POD modes, in order to highlight the most energetic structures. From 
Figure \ref{fig:structures_0}, long streak-like structures are clearly visible, 
typical of wall-bounded turbulence. Comparing this with Figure \ref{fig:structures_30}, it is apparent
that the long structures are indeed significantly less pronounced due to the stable stratification, 
potentially explaining the lack of the low-frequency mode in the swirl-switching as the thermal transient passes.

\section{Conclusion}\label{sec:conclusion}
In this paper, we studied the flow through a
U-bend that is subject to a thermal transient. An interesting feature is the
reversal of the secondary flow due to buoyancy. The thermal inertia of the wall
causes the near-wall fluid to sink, opposing the motion due to Dean vortices,
leading to a reversal of the secondary flow. Misalignment of the pressure
gradient and density gradient generates a baroclinic torque, which is the
dominant production term in the vorticity transport equation.

The impact of the secondary flow reversal on swirl-switching has been assessed.
It was found that the swirl number has a diminished amplitude for a period
while the reversal is establishing. Subsequent to this, there appears to be a
single dominant frequency for the reversed state, contrary to the isothermal
case in which there are two dominant frequencies (a low frequency mode,
superimposed upon a high-frequency mode). A physical explanation for this has
been proposed: the stable stratification due to the thermal transient leads to
lower levels of turbulent fluctuations in the upstream vertical pipe.
Low-frequency modes due to the large scale structures originating upstream of
the bend are therefore less pronounced. 

In the near-horizontal section of the pipe, we observe strong flow
stratification, and a large recirculation region over a portion of the
transient. The impact of the secondary-flow reversal on the heat and mass
transfer within the near-horizontal section is significant. One mode of heat
transfer into the stratification region is by the reversed secondary flow. This
highlights the importance of including the solid domain in this calculation;
without accounting for the thermal inertia of the wall, the secondary flow
would be driven by the centripetal force. Another mode of heat transfer into
the stratification region is via conduction through the solid pipe wall. Again,
this could not be accounted for without including the solid domain. 
We have also performed this computation with an adiabatic thermal boundary condition, 
in order to quantify the impact neglecting the solid domain has. Significant differences in temperature profiles between the conjugate and adiabatic cases have been observed. 

Although we have looked at flow in a U-bend, buoyancy-driven secondary flow
should be expected to occur in any mixed convection flows where the thermal
inertia of the wall is large, and a thermal transient of sufficient magnitude
propagates through the domain, including in straight pipes or ducts. A useful extension to this work may be to experimentally observe this reversal of the secondary flow under a broad parameter space.

\section*{Supplementary material}
Full 3D volumetric datasets containing ensemble averaged data can be downloaded from
\cite{skillen2019data}.

\section*{Acknowledgements}
This work was supported by the STFC Hartree Centre's Innovation Return on
Research programme, funded by the Department for Business, Energy \& Industrial
Strategy. Computational time on ARCHER, the UK National Supercomputing Service
(http://www.archer.ac.uk), was provided through the UK Turbulence Consortium,
grant number EP/R029326/1.
We are grateful to Dr. Adam Flint and the anonymous reviewers for their helpful advice and comments.

\section*{References}

\bibliography{ref}

\pagebreak

\section*{Nomenclature}
\begin{longtable*}{ p{0.3\textwidth} p{0.7\textwidth} }
    $D$           & Pipe inner diameter. \\
    $Dn$          & Dean number. \\
    $f$           & Frequency.   \\
    $g$           & Acceleration due to gravity.\\
    $R_c$         & Bend radius of curvature. \\
    $Ri$          & Gravitational Richardson number. \\
    $Ri_c$        & Centrifugal Richardson number. \\
    $Re$          & Reynolds number. \\
    $p$           & Pressure.\\
    $Pr$          & Prandtl number. \\
    $Pr_t$        & Turbulent Prandtl number. \\
    $Sw$          & Swirl number.\\
    $T$           & Temperature.\\
    $T_0$         & Reference (initial) temperature.\\
    $T_1$         & Final temperature.\\
    $t$           & Time.\\
    $\tilde{t}$   & Dimensionless time.\\
    $u$           & Velocity vector.\\
    $U$           & Bulk velocity.\\
    $\mathbf{x}$  & Position vector.\\
                  &                   \\
    $\alpha$      & Thermal diffusivity.\\
    $\beta$       & Thermal expansion coefficient.\\
    $\delta_{ij}$ & Kronecker delta.\\
    $\Delta$      & Filter width.   \\
    $\epsilon_{ijk}$ & Levi-Civita symbol.\\
    $\eta$        & Wall-normal direction.\\
    $\theta$      & Angle in cylindrical coordinate system. \\
    $\kappa$      & thermal conductivity.\\
    $\nu$         & Kinematic viscosity.\\
    $\rho$        & Density.\\
    $\rho_0$      & Reference density.\\
    $\omega$      & Vorticity vector.\\
\end{longtable*}

\pagebreak

\section*{Figures and tables (to be inserted at typesetting stage)}

\begin{table} 
\begin{tabular}{c c c}

Ensemble Size (runs) & $<T>_{RMS}$  & $<u^\prime u^\prime>_{RMS}$     \\ \hline
10 & 0.831 & 0.0519  \\ 
20 & 0.830 & 0.0548 \\  
30 & 0.830 & 0.0557 \\  
40 & 0.830 & 0.0559 \\  
\end{tabular}
\caption{Table showing convergence of root-mean-squared (RMS) $<T>$ and $<u^\prime
u^\prime>$ for different ensemble sizes. RMS values were integrated over the symmetry plane.}\label{tab:convergence}
\end{table}

\end{document}